\documentstyle[12pt,aaspp4,psfig,natbib]{article}

\newcommand{\vy}[2]{#1_{\scriptscriptstyle #2}}
\newcommand{\Ly}{Ly$\alpha$}

\def\gtorder{\mathrel{\raise.3ex\hbox{$>$}\mkern-14mu
             \lower0.6ex\hbox{$\sim$}}}
\def\ltorder{\mathrel{\raise.3ex\hbox{$<$}\mkern-14mu
             \lower0.6ex\hbox{$\sim$}}}
\def\proptwid{\mathrel{\raise.3ex\hbox{$\propto$}\mkern-14mu
             \lower0.6ex\hbox{$\sim$}}}
\def\arcsec{\ifmmode '' \else $''$\fi}

\def\arcsecpoint{\ifmmode ''\!. \else $''\!.$\fi}

\def\kms{\ifmmode {\rm km\ s}^{-1} \else km s$^{-1}$\fi}
\def\Msun{\ifmmode {\rm M}_{\odot} \else M$_{\odot}$\fi}
\def\Lsun{\ifmmode {\rm L}_{\odot} \else L$_{\odot}$\fi}
\def\Zsun{\ifmmode {\rm Z}_{\odot} \else Z$_{\odot}$\fi}

\def\ergscm2{ergs\,s$^{-1}$\,cm$^{-2}$}
\def\icm3{{\rm cm}^{-3}}
\def\icm2{{\rm cm}^{-2}}
\def\qo{\ifmmode q_{\rm o} \else $q_{\rm o}$\fi}
\def\Ho{\ifmmode H_{\rm o} \else $H_{\rm o}$\fi}
\def\ho{\ifmmode h_{\rm o} \else $h_{\rm o}$\fi}

\def\vFWHM{\ifmmode v_{\mbox{\tiny FWHM}} \else
            $v_{\mbox{\tiny FWHM}}$\fi}
\def\CCF{\ifmmode F_{\it CCF} \else $F_{\it CCF}$\fi}
\def\ACF{\ifmmode F_{\it ACF} \else $F_{\it ACF}$\fi}
\def\Halpha{\ifmmode {\rm H}\alpha \else H$\alpha$\fi}
\def\Hbeta{\ifmmode {\rm H}\beta \else H$\beta$\fi}
\def\Hgamma{\ifmmode {\rm H}\gamma \else H$\gamma$\fi}
\def\Hdelta{\ifmmode {\rm H}\delta \else H$\delta$\fi}
\def\Lya{\ifmmode {\rm Ly}\alpha \else Ly$\alpha$\fi}
\def\Lyb{\ifmmode {\rm Ly}\beta \else Ly$\beta$\fi}
\def\Lyg{\ifmmode {\rm Ly}\beta \else Ly$\gamma$\fi}
\def\hi{H\,{\sc i}}
\def\hii{H\,{\sc ii}}
\def\hei{He\,{\sc i}}
\def\heii{He\,{\sc ii}}
\def\heiii{He\,{\sc iii}}

\def\ciii{\ifmmode {\rm C}\,{\sc iii} \else C\,{\sc iii}\fi}
\def\civ{\ifmmode {\rm C}\,{\sc iv} \else C\,{\sc iv}\fi}

\def\nv{N\,{\sc v}}

\def\o5007{[O\,{\sc iii}]\,$\lambda5007$}

\def\ovi{O\,{\sc vi}}

\def\mgi{Mg\,{\sc i}}

\def\mgii{Mg\,{\sc ii}}

\def\siiv{Si\,{\sc iv}}

\def\siv{S\,{\sc iv}}

\def\caii{Ca\,{\sc ii}}
\def\fei{Fe\,{\sc i}}
\def\feii{Fe\,{\sc ii}}

\def\alii{Al\,{\sc ii}}
\def\aliii{Al\,{\sc iii}}

\def\o{\o}
\begin{document}
\title{THE INTRINSIC ABSORBER IN QSO~2359--1241: \\
 KECK AND HST OBSERVATIONS }

\author{Nahum Arav\footnote{Physics Department, University of
California, Davis, CA 95616 I: arav@astro.berkeley.edu}, 
Michael
S. Brotherton\footnote{ Kitt Peak National Observatory,  950 North Cherry Avenue, P. O. Box
26732, Tuscon, AZ 85726}\footnote{IGPP LLNL,
L-413, P.O. Box 808, Livermore, CA 94550}, 
Robert H. Becker$^{1,3}$,
Michael D. Gregg$^{1,3}$,
\\ Richard L.  White\footnote{Space Telescope Science Institute,
Baltimore, MD 21218}, Trevor Price$^{1,3}$, Warren Hack$^{4}$ }


\begin{abstract}          

We present detailed analyses of the absorption spectrum seen in
QSO~2359--1241 (NVSS J235953$-$124148).  Keck HIRES data reveal
absorption from twenty transitions arising from: \hei, \mgi, \mgii,
\caii, and \feii.  HST data show broad absorption lines (BALs) from
\aliii~$\lambda$1857, \civ~$\lambda$1549, \siiv~$\lambda$1397, and
\nv~$\lambda$1240.  Absorption from excited \feii\ states constrains
the temperature of the absorber to $2000\ltorder T \ltorder10,000$~K
and puts a lower limit of $10^5$~cm$^{-3}$ on the electron number
density.  Saturation diagnostics show that the real column densities
of \hei\ and \feii\ can be determined, allowing to derive meaningful
constraints on the ionization equilibrium and abundances in the flow.
The ionization parameter is constrained by the iron, helium and
magnesium data to $-3.0\ltorder \log(U) \ltorder-2.5$ and the observed
column densities can be reproduced without assuming departure from
solar abundances.  From comparison of the \hei\ and \feii\ absorption
features we infer that the outflow seen in QSO~2359--1241 is not
shielded by a hydrogen ionization front and therefore that the
existence of low-ionization species in the outflow (e.g., \mgii,
\aliii, \feii) does not necessitate the existence of such a front.  We
find that the velocity width of the absorption systematically
increases as a function of ionization and to a lesser extent with
abundance.  Complementary analyses of the radio and polarization
properties of the object are discussed in a companion paper
(Brotherton et al. 2000).

{\it Subject headings:} quasars: absorption lines

\end{abstract}          
\newpage
\section{INTRODUCTION}

The radio source NVSS J235953-124148 (z=0.868), hereafter
QSO~2359--1241, is unique among quasars.  It shows intrinsic
absorption from: \mgii\ and \feii, which appear in less than 1\% of
optically selected quasars; \mgi, which is even less frequent; and
from a meta-stable \hei\ level, which is only seen in two or three
other AGNs.  We use the term ``intrinsic absorption'' following the
definition given by Hamann et al (1997) and Barlow (1997). Whenever
this absorption is significantly blue-shifted with respect to the
systemic redshift of the quasar, we interpret it as rising from an
outflow connected with the AGN.  In \S~2 we establish the intrinsic
nature of the absorption seen in QSO~2359--1241.  Besides its rare
absorption features, QSO~2359--1241 has very high intrinsic
polarization ($\sim$5\%) and is moderately reddened (Brotherton et
al. 2000).  A low-resolution Keck spectrum of the object is shown in
figure 1.

By exhibiting absorption from \mgii\ and \aliii\ QSO~2359--1241 is
classified as a low-ionization BALQSO.  Low-ionization BALQSOs were
studied by Boroson and Meyers (1992), Voit, Weymann and Korista (1993)
Wampler, Chugai \& Petitjean (1995); Becker et al. (1997) and de Kool
et al. (2000), among others.  All the low-ionization BALQSOs show
absorption from \mgii, but only a subset of these show absorption from
\feii\ (Prominent examples include: QSO~0059--2735, Wampler et
al. 1995; Arp 102B, Halpern et al 1996; FIRST~0840+3633 and
1556+3517, Becker et al. 1997; QSO~1044+3656, de Kool et
al. 2000). An even smaller subset shows \mgi\ in absorption (Arp 102B,
QSO~1044+3656). For example, from the six objects studied by Voit,
Weymann and Korista (1993) only one (QSO~0059--2735) shows \feii\
absorption and none show \mgi\ absorption. QSO~2359--1241 shows
absorption features from all these ions as well as from \hei.  In the
discussion we elaborate on the conditions needed for detecting these
lines (both physical and observational) and argue that even though the
low-ionization features in QSO~2359--1241 are too narrow to be
classified as classical BALs, they are definitely part of a BAL
outflow because of their association with the much wider
high-ionization absorption troughs (\civ~$\lambda$1549 and
\siiv~$\lambda$1397) seen in the ultraviolet spectrum.

The ions detected in absorption and their unique characteristics make
QSO~2359--1241 a promising probe for the study of quasar
outflows. Features that can be used as diagnostics include:

\noindent 1) Appearance of relatively unblended absorption features
from both components of the \mgii\ doublet allows us to determine
whether the flow completely covers the emission region.  A partial
covering of the source is taken as a direct evidence for the intrinsic
nature of the absorber (Barlow 1997; Hamann et al. 1997; Arav et
al. 1999b), that is, an outflow associated with the AGN.  It also
shows that the apparent column densities extracted from the absorption
trough are only lower limits (Arav 1997; Arav et al. 1999b).

\noindent 2) Neutral helium absorption lines from the highly
meta-stable level 2$^3$S.  Very few AGNs show these lines, where known
examples are Mrk 231 (Boksenberg et al 1977; Rudy, Stocke \& Foltz
1985), NGC 4151 (Anderson 1974) and perhaps 3CR 68.1 (Brotherton et
al. 1998).  The \hei\ lines are important diagnostics for the
ionization state of the gas.  From the ionization equilibrium of this
level we can infer lower limits on the He$^+$ column density and
therefore lower limits on the \hii\ column density.  The different
oscillator strengths of the detected lines allow us to determine the
real optical depth and covering factor of the \hei\ absorbers.

\noindent 3) Detection of absorption from excited states of \feii\
allows for lower limits and sometimes even determination of the number
density of electrons ($n_e$) in the gas. In a study of QSO~1044+3656
(de Kool et al.  2000) we were able to determine $n_e$ in the outflow
and combined with photo-ionization constraints showed that the outflow
is situated about 1000 pc. from the central source.

\noindent 4) Appearance of \mgi\ absorption necessitates a low
ionization parameter, since it is  difficult to shield \mgi\ from
ionizing photons. (The ionization parameter $U$ is defined as the
ratio of number densities between hydrogen ionizing photons and hydrogen
nuclei in all forms.)

To realize this diagnostic potential we observed the optical spectrum
of QSO~2359--1241 using the HIRES spectrograph on the Keck telescope.
In this paper we concentrate on a detailed analysis of the numerous
absorption features seen in our high-resolution ground-based
spectroscopic data (\S~2), and in low-resolution HST UV prism data
(\S~3).  In \S~4 we analyze the ionization equilibrium and abundances
of the outflow.  The discovery of the object, its radio, polarization,
and overall optical characteristics are described in a companion paper
(Brotherton et al. 2000).

\begin{figure}
\centerline{\psfig{file=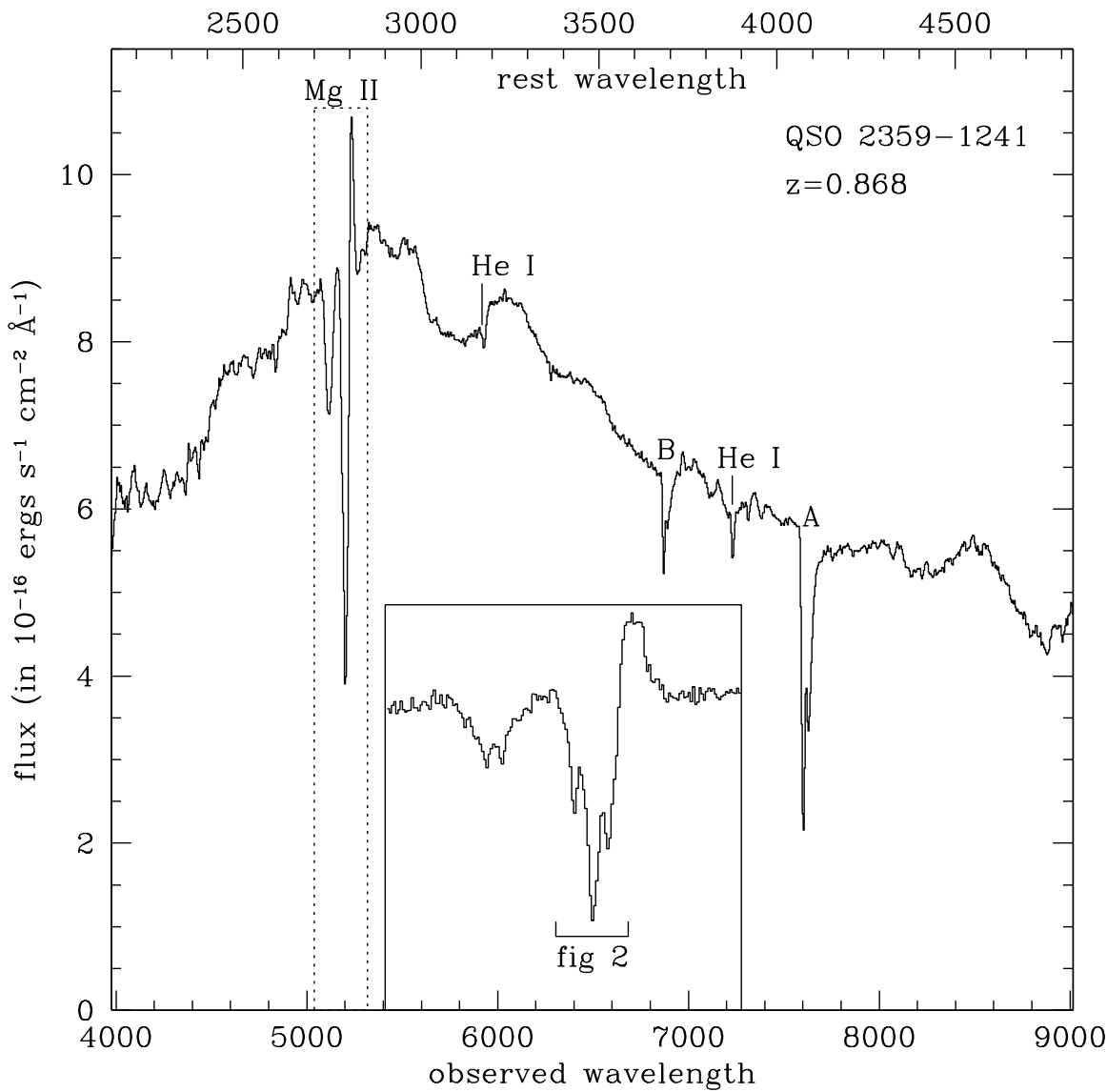,height=16.0cm,width=16.0cm}}
\vspace{-1cm}
\caption{Low resolution Keck spectrum with major absorption features
marked.  An expanded view of the \mgii\ region is shown in the insert,
which covers the spectral interval bounded by the dotted rectangle.
The data shown in the insert is from a somewhat higher resolution
observation done at Lick. For comparison, figure 2 shows Keck-HIRES
data covering only the marked spectral region within the insert.
}\label{low_res}
\end{figure}

\newpage

\begin{figure}
\vspace{-3cm}
\centerline{\psfig{file=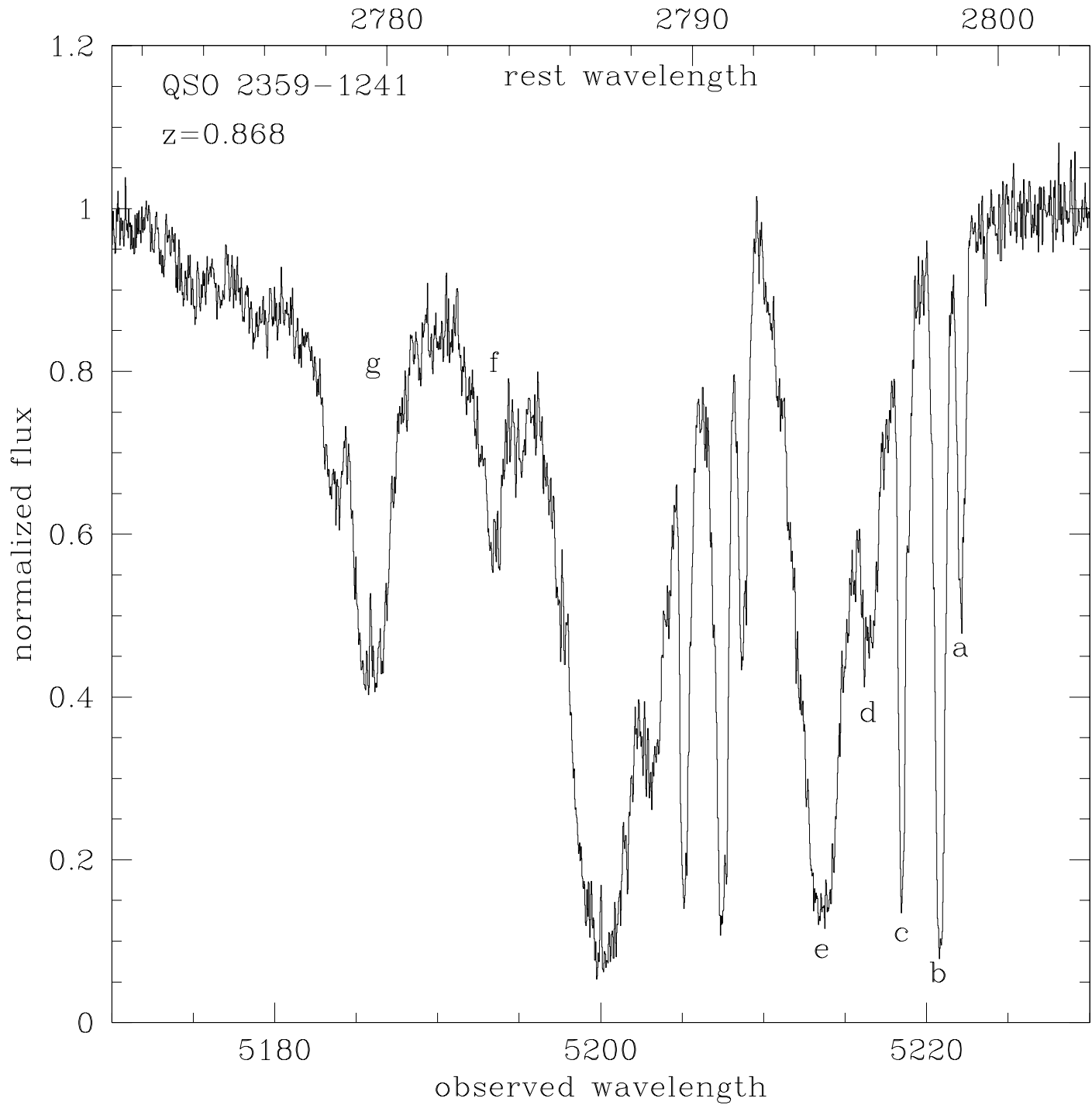,height=20.0cm,width=20.0cm}}
\vspace{-1cm}
\caption{Normalized HIRES data (see text) of the \mgii\ absorber.
Individual features are marked: $a-e$ in the red doublet component,
$f$ and $g$ in the blue doublet component. The spectral coverage of
this plot is marked in the insert to Fig. 1.
}\label{hires_mg2}
\end{figure}

\section{INTRINSIC ABSORPTION IN THE KECK HIRES SPECTRUM}

\subsection{Data Acquisition and Reduction}

On December 26, 1998 we used the High Resolution Echelle Spectrometer
(HIRES, Vogt et al.\ 1994) on the Keck~I 10-m telescope to obtain $4
\times 1500$ second exposures of QSO 2359$-$1241 covering 4330 -
7450\AA\ using a 1\farcs1 wide slit.  The orders overlap up to
6410\AA, beyond which small gaps occur between orders.  The slit was
rotated to the parallactic angle to minimize losses due to
differential atmospheric refraction.  The observing conditions were
excellent with sub-arcsecond seeing and near-photometric skies.  The
spectra were extracted using routines tailored for HIRES (Barlow
2000).  The resolution of the final spectrum 
is $R=39000$.

A smooth continuum was fit to the spectrum in regions free of
absorption or emission features.  This procedure is somewhat
subjective, particularly in the case of blended emission lines (such
as \feii).  While unlikely to create any false
absorption features, this normalization introduces some uncertainty in
the true continuum level; consequently, equivalent widths of
absorption features are somewhat uncertain.

\subsection{Intrinsic Nature of the Absorption}

Many of the absorption features in the HIRES data arise from a complex
intrinsic absorber.  The evidence for the absorber being intrinsic is
based on: 1) Comparing the absorption features seen in the \mgii\
doublet (the apparent optical depth ratio differs from the nominal
value of 1:2).  2) Existence of \hei\ absorption from a meta-stable
state, which requires the density of the absorber to be several orders
of magnitude larger than is seen in the ISM and IGM.  3) Existence of
\feii\ absorption from excited levels, which also necessitates high
density.  4) Appearance of full fledge BALs (\civ~$\lambda$1549,
\siiv~$\lambda$1397) in the HST FOC spectrum which coincide in
velocity with the absorption seen in the HIRES data.  Each of these
separate pieces of evidence is enough by itself to identify the
absorption system as intrinsic.  In the rest of this section we
analyze the features in the HIRES data that are associated with this
intrinsic absorber.

\subsection{\mgii}

A high resolution spectrum of the \mgii\ absorption feature reveals a
rich and complex structure.  In figure 2 we show HIRES data for 
a part of the \mgii\ absorption, the spectral region shown in figure
2 is marked on the insert to figure 1.  We have labeled five distinct
absorption features ($a-e$) associated with the red doublet component, 
which are also seen in the blue doublet component.  Absorption
features $f$ and $g$ are seen in the blue doublet component, but
their red counterparts are blended with the ($a-e$) blue complex.

The appearance of the same features in both doublet components allow
us to study the covering factor and real optical depth of the outflow.
However, the situation in QSO~2359--1241 is not ideal since the red
component of feature $g$ falls in the middle of blue component of
feature $e$; the red component of feature $f$ contaminates the blue
component of feature $b$; and there is a shallower red absorption
contribution across the entire blue absorption structure $a-e$. These
contaminations do not allow for a clean simultaneous solution for the
covering factor and the optical depth (as was done for the \siiv\ BAL
in QSO 1603+3002; Arav et al 1999b).  However, most of the
contaminating absorption is rather shallow and we can still get
semi-quantitative results from analyzing the relationship between the
same absorption features seen in both doublet components.

To analyze the residual intensities we first need to normalize the
data.  Since the absorption occur on the blue wing of the \mgii\ broad
emission line (BEL), there are two physical choices here.  First, the
outflow may cover both the continuum source and the BEL region. In
this case we model the emission line with two Gaussians and divide the
data by the assumed unabsorbed emission, which consists of the BEL
plus continuum.  Second, the outflow may only cover the continuum
source and not the BEL region.  In BALQSO~1603+3002 there is strong
evidence that the flow does not cover the BELs (Arav et al 1999b).
For this case we need to subtract the modeled BEL from the data in
order to determine the flux seen by the absorber.  We then divide this
flux by the continuum to obtain the normalized flux seen by the
absorber.  We will argue below that the second scenario is more
probable and use this normalization in figure \ref{hires_mg2}.

\begin{figure}
\vspace{-3cm}
\centerline{\psfig{file=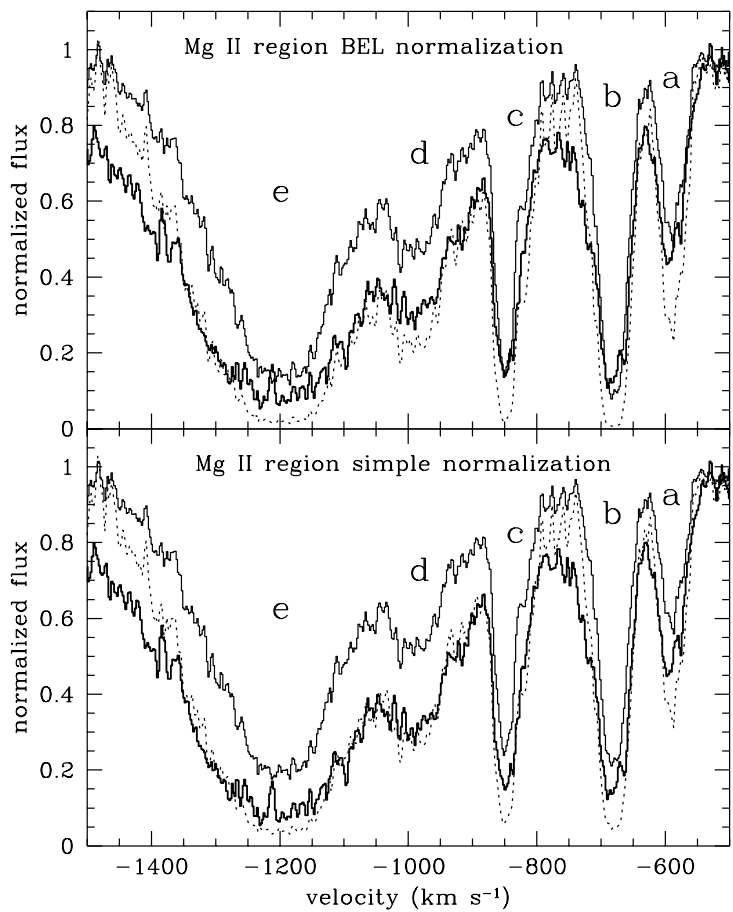,height=14.0cm,width=14.0cm}}
\vspace{-1cm}
\caption{\mgii\ absorption data in two different normalizations. Bottom
panel shows simple normalization, data divided by the sum of modeled continuum
and modeled BEL.  The thick solid line shows the data for the blue doublet
component; the thin solid line indicates the red doublet
component; the dotted line shows the expected blue doublet absorption
given the red component data.  Top panel shows the same for the
case where we assume the absorption covers the continuum region but not
the BEL region.  This is done by subtracting a modeled BEL from the
data and dividing the result by the continuum.  
}\label{mg2_abs_fits_combined}
\end{figure}

Figure \ref{mg2_abs_fits_combined} shows a comparison of the \mgii\
absorption features in each doublet component.  For direct comparison
we use velocity presentation and the two panels show the two
normalizations discussed above.  We use a thick solid line for the
blue component data and a thin line for the red one.  If the flow
fully covers the QSO's emission (both continuum source and BEL
regions), the expected residual intensity of the absorption features
seen in the blue doublet component is:
$I_b(v)$[expected]$=I^2_r(v)$[observed]; where $I_r$ is the residual
intensity of the absorption features seen in the red doublet
component.  We plot this expected blue residual intensity in figure
\ref{mg2_abs_fits_combined} as a dotted line.  In both normalizations
it is evident that the depth of features $a, b$ and $c$ in the blue
doublet component is smaller than expected by assuming complete
coverage.  This is a clear indication for partial covering and hence
for the intrinsic nature of the absorption (Barlow 1997). We note that
partial covering is not restricted only to geometrical coverage.  The
photons at the bottom of the troughs may also arise from scattering
contribution.  If we assume that the flow does not cover the BEL
region, the top panel in figure \ref{mg2_abs_fits_combined} show that
the residual intensities for features $b$ and $c$ are identical within
the errors in both doublet components.  In this case both features are
saturated ($\tau_{red}\gtorder3$) and the shape of the absorption
profile is completely dependent on the covering factor.  If the flow
covers the continuum and the BEL equally, then from the bottom panel
of figure \ref{mg2_abs_fits_combined} we also infer that the blue
doublet absorption of features $a, b$ and $c$ is not deep enough if we
assume complete coverage.  Therefore, we conclude that no matter which
normalization we use the flow in features $a, b$ and $c$ show partial
coverage and hence demonstrate the intrinsic nature of the absorber.

Which of the two covering scenarios is more probable?  In the case of
BALQSO~1603+3002 we argued that ``in the absence of a physical preference
for $\tau_{real}$ values of order unity, values between 2--5
necessitate some fine tuning whereas the range $5-\infty$ is simply
much more probable numerically.''  This argument suggests that it is
more probable that features $b$ and $c$ do not cover the BEL region.
However, we notice that the equal covering normalization (bottom panel
in Fig.  \ref{mg2_abs_fits_combined}) shows that feature $d$ might
completely cover both the continuum source and the BEL region
(expected and observed blue component are equal).  This scenario is
simpler, since it does not necessitate a partial covering factor, and
thus is more appealing.  We suggest that the first argument is somewhat
stronger and therefore it is more probable that features $b$ and $c$
do not cover the BEL region.  However, the evidence is weaker than is
seen in BALQSO~1603+3002.

In addition to components $a-g$ there is a high velocity trough at
$-5000$ \kms, which can be seen in the insert to figure 1.  We do not
show the HIRES data since they do not reveal qualitatively different
structure than the one seen in the insert.  A \mgii\ doublet structure
in evident in the feature and the HIRES data confirm that the depth
ratio of the two components is 1:1, that is, complete saturation.
This 1:1 depth ratio is caused by partial covering factor, which
indicates that the absorption system is intrinsic.


\subsection{Apparent Column Densities}

In Table 1 we give the apparent column density measurements of the
intrinsic absorption features for all the identified lines.  
Apparent column density are derived by substituting 
the apparent optical depths of the troughs 
(defined as $\tau=-ln(I_r)$, where
$I_r$ is the normalized residual intensity seen in the trough) in:
\begin{equation} \rm{N_{ion}} = 
 \frac{ 3.7679 \times 10^{14}~\rm{cm}^{-2} }{
\lambda_{\rm{o}}\rm{f_{ik}} } \int {\tau\/(v) dv}, 
\label{eq:column}
\end{equation} 
where $\lambda_{\rm{o}}$ and $\rm{f_{ik}}$ are the transition's
wavelength and oscillator strength, and where the velocity is measured
in \kms.  We note that the apparent column densities give good
estimates for the real column densities only if the absorbing material
covers the emission source completely and uniformly and where
scattered-photons do not contribute appreciably to the residual
intensity in the troughs (see discussions in Korista et al 1992; Arav
1997).  Otherwise, apparent column densities are only lower limits on
the real ones (Arav et al 1999a, Arav et al 1999b ). We define
$gf\lambda$ (where $g$ is the statistical weight, $f$ is the
oscillator strength and $\lambda$ is the transition's wavelengths) as
the strength factor of the line and give its value (normalized to that
of the strongest transition; data from Verner, Verner \& Ferland
1996).  For the \feii\ lines we multiplied the expression in equation
(\ref{eq:column}) by the ratio of  statistical weight of the lower level
(summed over all observed states) to that of the specific state.  This
procedure yields independent estimates for $\rm{N_{\feii}}$ from each
transition provided the level populations are in  LTE.(see \S~2.6).  
In \mgii\ components $c$ and $d$ are
embedded in a wide extension of component $e$, their integration
interval only covers the regions where these feature are distinctive,
without compensation for the fact that some of the absorption is due
to the wing of component $e$. For the other lines the different
absorption components are well separated, and therefore the
integration over the components is straightforward. For the \hei\ measurments
 we corrected for the atmospheric absorption seen in the spectral
region around the intrinsic \hei$\lambda3890$ line (especially evident
on the red side of component $e$, see Fig. \ref{hires_he1}). The estimated
error for each line is derived by performing the integration given in equation
(\ref{eq:column}) on two absorption free regions in the spectral
vicinity of the measured features and taking the average of their
absolute values.  This procedure takes into account both
signal-to-noise and continuum uncertainties.  The error is formally
derived for component $e$ and can be used as a conservative estimate
for the other components since they are narrower. For \mgii\ the error
is significantly larger since we had to take the blending of the
different components into account.


\begin{table}
\begin{center}
\begin{tabular}{llcccccccc}
\multicolumn{10}{c}{\sc Table 1: HIRES Apparent Ionic Column Densities} 
\\[0.2cm]
\hline
\hline
%
\multicolumn{1}{c}{Ion}
&\multicolumn{1}{c}{Transition}
&\multicolumn{1}{c}{Strength$^a$}
&\multicolumn{1}{c}{N(e)$^b$}
&\multicolumn{1}{c}{N(d)$^b$ }
&\multicolumn{1}{c}{N(c)$^b$ }
&\multicolumn{1}{c}{N(b)$^b$ }
&\multicolumn{1}{c}{N(a)$^b$ }
&\multicolumn{1}{c}{error$^c$ }
&\multicolumn{1}{c}{N(tot)$^b$ }
\\
\multicolumn{1}{c}{(1)} & \multicolumn{1}{c}{(2)} & 
\multicolumn{1}{c}{(3)} & \multicolumn{1}{c}{(4)} &
\multicolumn{1}{c}{(5)} & \multicolumn{1}{c}{(6)} & 
\multicolumn{1}{c}{(7)} & \multicolumn{1}{c}{(8)} &
\multicolumn{1}{c}{(9)} & \multicolumn{1}{c}{(10)}
\\[0.05cm]
\hline
\caii & 3970 & 1   & 12.43 & 11.06 & 11.68 & 11.59 & 10.90 & 11.1 & 12.57 \\
\hei  & 3889 & 1   & 13.80 & 12.70 & 13.14 & 13.19 & 12.76 & 12.3 & 14.02 \\
\hei  & 3189 & .36 & 13.85 & 12.29 & 12.22 & 12.94 & 12.49 & 12.8 & 13.90 \\ 
\hei  & 2946 & .17 & 13.87 & 12.47 & 12.96 & 13.30 & 12.82 & 13.0 & 14.00 \\ 
\mgi  & 2853 & 1   & 11.88 & 10.98 & 11.44 & 11.53 & 10.68 & 10.7 & 12.18 \\ 
\mgii & 2803 & 1$^d$ & 14.21 & 13.21 & 13.30 & 13.38 & 12.89 & 12.6 & 14.43$^e$ \\
\feii & 2632$^{*,f}$ & 0.46 & 13.59 & 12.84 & 11.45 & 12.25 & 12.32 & 12.6 & 13.65 \\ 
\feii & 2612$^*$     & 0.33 & 13.65 & 13.13 & 13.09 & 11.77 & 12.50 & 12.8 & 13.83 \\
\feii & 2608$^*$     & 0.22 & 13.83 & 12.35 & 13.02 & 12.87 & 13.10 & 13.1 & 14.00 \\ 
\feii & 2600         & 0.76 & 13.66 & 12.83 & 12.76 & 12.91 & 12.14 & 12.5 & 13.83 \\  
\feii & 2587         & 0.17 & 13.92 & 12.56 & 13.47 & 13.30 & 12.88 & 13.1 & 14.16 \\
\feii & 2405.6$^*$   & 0.43 & 13.63 & 12.37 & 12.74 & 13.00 & 12.01 & 13.0 & 13.61 \\ 
\feii & 2396.4$^*$   & 0.68 & 13.62 & 12.11 & 11.42 & 13.08 & 11.80 & 12.8 & 13.73 \\ 
\feii & 2382         & 1.00 & 13.61 & 12.67 & 12.71 & 12.87 & 12.75 & 12.6 & 13.81  
\\[0.01cm]
\hline
\end{tabular}
\end{center}
$^a$ - Expected absorption strength ratio for lines from the same ion 
(essentially normalized $gf\lambda$, see text).\\
$^b$ - Log$_{10}$ of the column density for each absorption 
subcomponent (see Figs. 4 and 5). \\
$^c$ - Also in units Log$_{10}$ of the column density.
The error is roughly appropriate for each individual component, and 
 takes into account both signal-to-noise and continuum uncertainties. \\ 
$^d$ - Although the red doublet component of \mgii\ is only as half as strong 
as the blue one, we give it strength=1 since 
we do not report the blue component measurements separately.   \\
$^e$ - Total column density for \mgii\ includes contributions from component
$f$ ($\log(N)=13.41$) and $g$ ($\log(N)=13.06$), which are associated with the 
blue component of the \mgii\ doublet. \\
$^*$ - Transition from excited level. \\
$^f$ - A blend of two excited transitions \feii\ $\lambda$2632.11 and 
\feii\ $\lambda$2631.83. \\
\end{table}

\pagebreak

\subsection{\hei }

\begin{figure}
\vspace{-3cm}
\centerline{\psfig{file=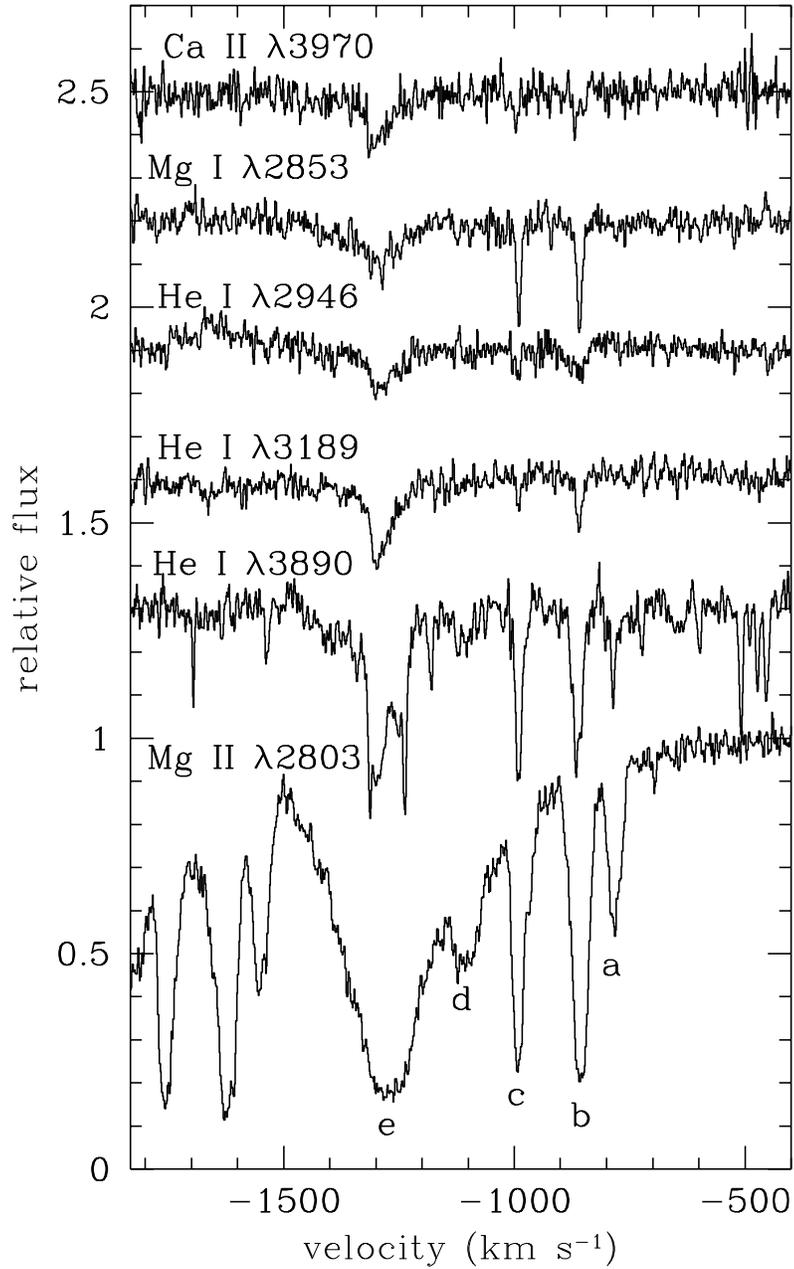,height=24.0cm,width=18.0cm}}
\vspace{-3cm}
\caption{Absorption troughs from three \hei\ lines,
\caii~$\lambda$3970 and \mgi~$\lambda$2853 are plotted on the same
velocity scale together with the intrinsic absorption seen in \mgii.
The sub-troughs are labeled underneath the \mgii\ data. 
}\label{hires_he1}
\end{figure}

Figure \ref{hires_he1} shows absorption in \hei, \mgi\ and \caii\
associated with the intrinsic absorber.  All the lines are plotted on
the same velocity scale where the wavelength of the transition in the
rest frame of the object is at 0 \kms.  The spectral segments are
plotted in the same normalized flux scale which is shifted for each
line for presentation purposes.  Absorption features from the He
triplet lines are easily recognized as being part of the intrinsic
outflow.  The column densities for the three \hei\ lines in each
feature are in agreement given the errors, implying that unlike the
\mgii\ case the \hei\ absorption is not saturated.  This finding is a
prerequisite for determining the ionization equilibrium and
abundances in the flow (see \S~4), since it gives us the actual column density
as opposed to a lower limit available from the apparent column density.

The observed \hei\ lines all arise from the meta-stable level
2$^3$S. Extensive treatment of this level appears in the literature
and the following discussion is largely based on: Macalpine (1976);
Rudy, Stocke, \& Foltz (1985); Clegg (1987); Oudmaijer, Busfield, \&
Drew (1997), all of which rely to some extent on Osterbrock's
``Astrophysics of Gaseous Nebulae'' (1974).  In equilibrium, the
population of the 2$^3$S is determined by the balance of arrivals from
recombination to all triplet levels versus departures mainly due to
collisional transition to other levels.  Since under most conditions
all recombinations to the triplet levels end up in the 2$^3$S level we
obtain:

\begin{equation} 
\vy{n}{He^+}n_e\vy{\alpha}{T}=\vy{n}{2^3S}\left[A_{21}+n_e(q_{tr}+q_{ci})
+\int_{\vy{\nu}{0}}^{\infty}\frac{a_{\nu}L_{\nu}}{4\pi r^2h\nu}d_{\nu}\right],
\label{eq:he}
\end{equation} 
where $\vy{n}{He^+}$ is the number density of singly ionized helium,
$n_e$ is the electron number density $\vy{\alpha}{T}$ is the total
recombination coefficient to all triplet levels, $\vy{n}{2^3S}$ is the
number density of neutral helium in the 2$^3$S level, $A_{21}$ is the
Einstein A coefficient for the forbidden transition (625 \AA) from the
2$^3$S level to the ground level (1$^1$S), $q_{tr}$ is the rate of
collisional transfer to all singlet level (which is dominated by
collisions to the 2$^1$S and 2$^1$P levels), $q_{ci}$ is the
collisional ionization rate which becomes important above 20,000 K
(Clegg 1987), $a_{\nu}$ is the photoionization cross section for
2$^3$S, $L_{\nu}/(4\pi r^2h\nu)$ is the flux of ionizing photons
($L_{\nu}$ is the luminosity per unit frequency, $r$ is the distance
to the emitting source and $h$ is Planck's constant), $\vy{\nu}{0}$ is
the threshold frequency for ionizing the 2$^3$S level (4.77 eV, 2600
\AA).

Equation (\ref{eq:he}) can give us a maximum for the
$\vy{n}{He^+}/\vy{n}{2^3S}$ ratio.  Neglecting photoionization,
collisional ionization and radiative transition to the ground level
(i.e., assuming $n_e$ larger than the critical density of
$3\times10^3$~cm$^{-3}$) we obtain:
$\vy{n}{He^+}/\vy{n}{2^3S}=\vy{\alpha}{T}/q_{tr}$ Using the values
given for $\vy{\alpha}{T}$ and $q_{tr}$ given in Clegg (1987) and
Osterbrock (1974) we find $\vy{n}{2^3S}/\vy{n}{He^+}=6\times10^{-6}$
More generally, Clegg (1987) gives the above ratio as function of
$n_e$ and temperature (including radiative transition to the ground
level but neglecting photoionization) as:
\begin{equation} 
\frac{\vy{n}{2^3S}}{\vy{n}{He^+}}=\frac{5.8\times10^{-6}T_4^{-1.19}}
{1+3110T_4^{-0.51}n_e^{-1}}
\label{eq:he_ratio}
\end{equation} 
where $T_4$ is the temperature in units of $10^4$ K.  Equation
(\ref{eq:he_ratio}) is a good approximation for $8,000<T<20,000$,
where in our case the temperature might be somewhat lower (see below).
We can use these estimates combined with our measurement for the total
column density seen in the \hei\ metastable lines, to set a minimal
He$^+$ column density of $\sim 2\times10^{19}$ cm$^{-2}$ in the
intrinsic absorber.  Assuming solar abundances this estimate yields a
minimal \hii\ column density of $\sim 2\times10^{20}$ cm$^{-2}$.


\pagebreak

\subsection{\feii}
\begin{figure}
\vspace{-3cm}
\centerline{\psfig{file=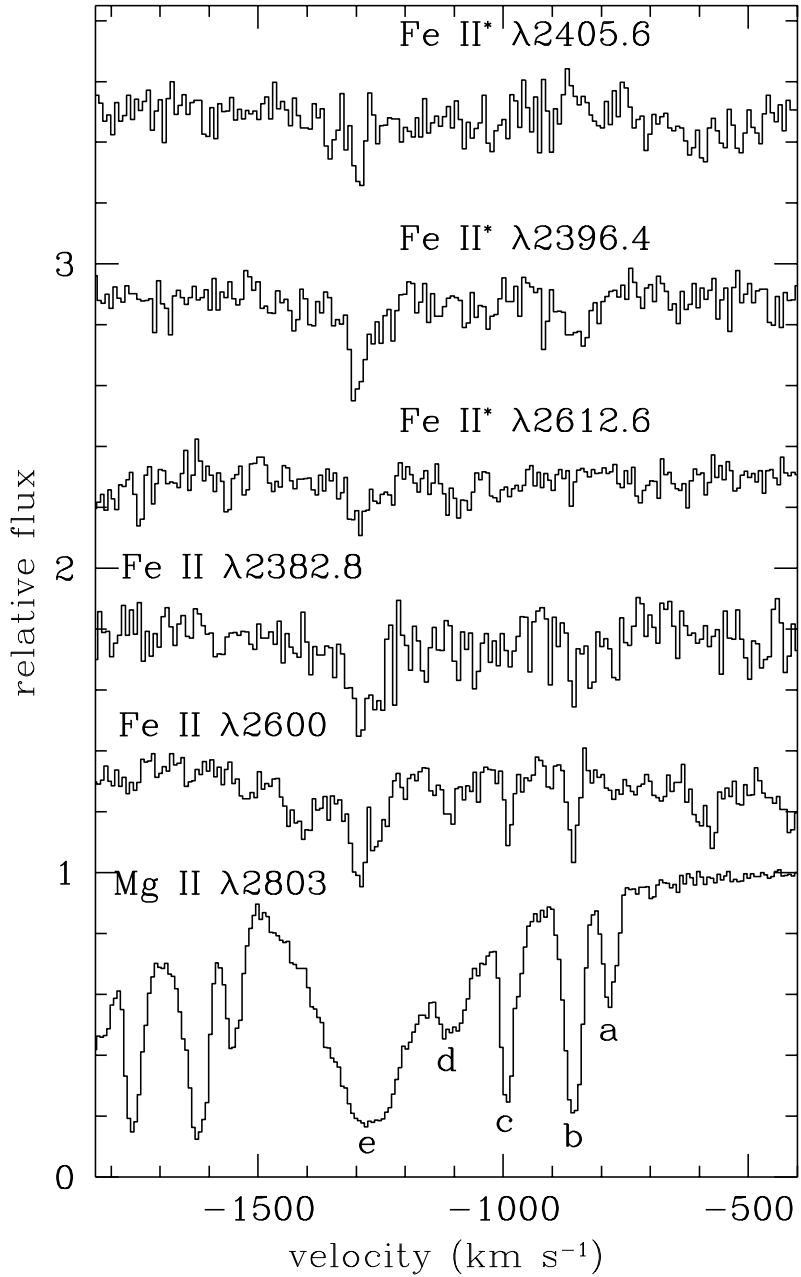,height=24.0cm,width=18.0cm}}
\vspace{-3cm}
\caption{Similar to Fig, \ref{hires_he1}  for five \feii\ transition,
two of which are from excited states. Data smoothed by four pixels to 
compenste for lower signal-to-noise at the blue end of the HIRES spectral coverage.
}\label{hires_fe2}
\end{figure}

Many \feii\ absorption features are detected in the HIRES data.  In
table 1 we give column density measurements for the most unambiguous
detections and in figure \ref{hires_fe2} we plot the absorption
associated with five of these on the same velocity scale of the
absorption seen in \mgii\ $\lambda$2803 line.  The S/N of the 
\feii\ features is lower than that of the magnesium and \hei\ features
due to the lower throughput of the HIRES detector at shorter wavelengths.
For this reason we concentrate our discussion on feature $e$ which is
the strongest one detected in all the lines.

Absorption features from excited levels of \feii\ are clearly detected.
These transitions arise from energy levels 0.05 eV. ($\lambda$2612,
$\lambda$2396.4) and 0.08 eV. ($\lambda$2608, $\lambda$2405.6) above
ground. Transitions from two slightly higher energy levels are also
detected (0.11 and 0.12 eV.), although we do not include them in the
table since their detection significance is lower.  The $\rm{N_{\feii}}$ 
values given in table 1 assume
LTE population of the
levels at the limit $kT\gg\Delta E$ (where $k$ is Boltzmann's
constant, $T$ is the temperature and $\Delta E$ is the energy
difference between the levels).  That is, the ratio of optical depth
of two features is given by:
\begin{equation} 
\frac{\tau_1}{\tau_2}=\frac{n_1f_1\lambda_1}{n_2f_2\lambda_2}=
\frac{g_1f_1\lambda_1}{g_2f_2\lambda_2},
\label{eq:tau_ratio}
\end{equation} 
where $n_1$ and $n_2$ are the level populations of lower level that
give rise to each transition; $\lambda_1$ and $\lambda_2$ are the
transitions' wavelengths, $f_1$ and $f_2$ are the oscillator strengths
and $g_1$ and $g_2$ are the statistical weights of the levels.  For
the last equality we used the assumption of LTE population at the
limit $kT\gg\Delta E$.  We define $gf\lambda$ as the strength
factor of the line and give its value (normalized to that of the strongest
transition; data from Verner, Verner \& Ferland 1996) in table 1.  
Six of the lines (from both ground and excited states) give consistent
estimates for $N_{\rm{\feii}}$, validating the LTE assumption, 
and like the \hei\ case show that 
the inferred $N_{\rm{\feii}}$ are actual determinations and not lower limits.
The two  inconsistent $N_{\rm{\feii}}$ estimates are probably due to uncertainties in the oscillator strength of these transitions (see de Kool et al. 2000).

Since the highest energy level we detect
(0.12 eV.) is equivalent to a temperature of $\sim1000$ K, we infer
that the absorbing gas is at $T\gtorder2000$~K.  Lower temperature will
cause a significant reduction in the higher level population due to the
exponential factor in the Boltzmann equation, and this is not seen in
the data.  We are also able to constrain the temperature from above
due to the non-detection of \feii\ $\lambda$2563, which arise from a
1.0 eV. level.  In order to supress the 1.0 eV. level population below
our detection limit we must have $T\ltorder10,000$~K.  The detection of
\feii\ excited transitions necessitates the gas to be above the
critcal density for these transitions.  From this we infer 
$n_e\gtorder 10^5$~cm$^{-3}$ (see de Kool
et al. 2000).  Finally, the non detection of \fei\ $\lambda2524$
necessitate $\log(U)>-5$.  Photoionization models (see \S~4) show that
 at smaller values of $U$ we should have detected
an appreciable absorption in the \fei\ line given the inferred \feii\
column density (see Fig. \ref{cloudy_u}).


\pagebreak

\subsection{ \mgi\ and \caii }

In figure \ref{hires_he1} we identify components $b, c$ and $e$ in
\mgi\ $\lambda2853$.  A detection of this line is significant since
the ionization potentiol of the ion is 7.6 eV.  As disscussed in de
Kool et al. (2000), this low ionization energy does not allow a
hydrogen ionization front to protect the \mgi\ ions from destruction by shielding
it from ionizing photon.  This is in contrast to the case of \mgii\
with ionization potentiol of 15.0 eV, which is higher than that of
hydrogen (13.6 eV). Therefore, a hydrogen ionization front can protect
the \mgii\ ions from photodissociation (Voit, Weymann \& Korista
1993).  Since shielding does not work for \mgi\, a low ionization
parameter is needed in order for it to survive in the typical
radiation environment produced by the quasar.  We discuss this issue
further in \S~4.

Components $b$ and $c$ are most prominenet in \mgi.  This gives
independant support to the conclusion from the \mgii\ analysis that
componets $b$ and $c$ have larger optical depth compared to the other
components.  In \mgi\ we do not detect components $a$ and $d$.  Using
the upper limits for $a$ and $d$ we conclude that $\tau_b$ and $\tau_c$
are at list five times larger than $\tau_a$ and $\tau_d$.
Our spectral coverage also contains the red  component of the 
\caii\ doublet ($\lambda$3970).  The line is clearly detected in 
component $e$ and marginally deteced in components $b$ and $c$.

\section{HST FOC SPECTRUM}
\begin{figure}
\centerline{\psfig{file=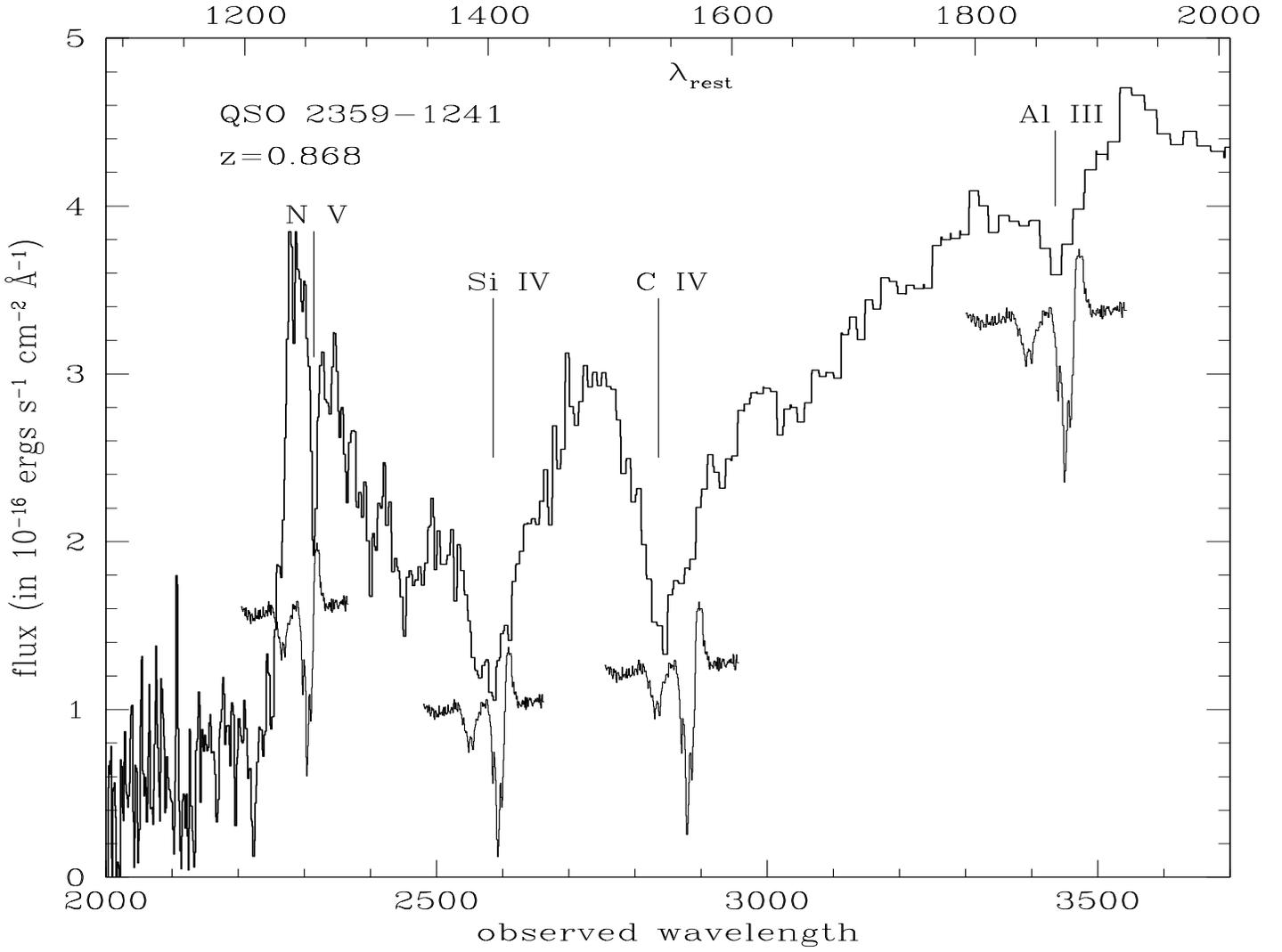,height=20.0cm,width=20.0cm}}
\vspace{-2cm}
\caption{Portion of the FOC spectrum with main absorption features
labeled.  These are identified using the \mgii\ ground-based template
(see text).  Good matches are seen for intrinsic absorption in:
\aliii~$\lambda$1857, \civ~$\lambda$1549, \siiv~$\lambda$1397 and
\nv~$\lambda$1240.
 }
\label{foc_linear_shift}
\end{figure}

A low-resolution UV spectrum of QSO~2359--1241 was obtained using the
Faint Object Camera (FOC) on-board the HST.  The data reduction and
characteristics are described in Brotherton et al (2000).  Figure
\ref{foc_linear_shift} shows part of the FOC spectrum.  
 Two deep and wide absorption features
are seen in the spectrum.  These are \civ~$\lambda$1549 and
\siiv~$\lambda$1397 BALs from the same system.  Several additional
absorption features are also noticeable.  We use the \mgii\ absorption
template to determine the relationship between the absorption features
seen in the FOC data to those seen in the ground-based observations.
To identify FOC absorption features associated with the \mgii\
absorption, we displace the \mgii\ template to the expected wavelength
position of a candidate transition.  This is done by multiplying the
rest wavelength of the template by
$\lambda_{c}/\vy{\lambda}{\mbox{\mgii}}$, where $\lambda_{c}$ is the
wavelength of the candidate transition.  The result (shown in figure
\ref{foc_linear_shift}) confirms the existence of absorption features
from the resonance lines \aliii~$\lambda$1857, \civ~$\lambda$1549,
\siiv~$\lambda$1397, and \nv~$\lambda$1240 associated with the \mgii\
intrinsic absorption.  We also note that the steep intensity drop on
the red wing of the \Ly\ BEL can be explained by a \Ly\ BAL from the
same system.

As evident from figure \ref{foc_linear_shift}, there are differences
in the shape of the absorption features in the FOC data to the \mgii\
template, especially the \civ\ feature.  Some of these are caused by
the much lower resolution of the FOC spectrum.  However, the fact that
the \siiv\ and \civ\ absorption features are more extensive than can
be extrapolated from the \mgii\ template is not unique.
QSO~0059--2735 shows a similar behavior (Weymann et al 1991, Wampler,
Chugai \& Petitjean 1995), where the \siiv\ and \civ\ BALs are much
wider than the \mgii\ BAL.  In table 2 we give apparent column
densities for the BALs seen in the FOC spectrum.  A comparison with
table 1 shows that the apparent column density of \aliii\ is similar
to that of \mgii, a phenomenon that is also observed in QSO~0059--2735
(Weymann et al. 1991, Wampler, Chugai \& Petitjean 1995) and in
QSO~1232+1325 (Voit, Weymann \& Korista 1993).  In contrast, the
apparent column density of the \civ\ and \siv\ BALs are roughly an
order of magnitude larger than the \mgii\ BAL, which is a consequence
of the much larger width of the \siiv\ and \civ\ BALs.  Again this is
similar to what is observed in QSOs 0059--2735 and 1232+1325. We
elaborate further on the relationship between the high and low
ionization absorbers in the discussion.

In summary, the FOC spectrum shows the following BALs associated with
the \mgii\ intrinsic absorption: \aliii~$\lambda$1857,
\civ~$\lambda$1549, \siiv~$\lambda$1397, and \nv~$\lambda$1240.  The
low resolution of the FOC spectrum does not allow for much analysis of
these BALs, other than the fact that the \civ~$\lambda$1549,
\siiv~$\lambda$1397 are full fledged BALs with a full-width-half-maximum
 of $\sim8000$
\kms.

\begin{table}[htb]
\begin{center}
\begin{tabular}{lcc}
\multicolumn{3}{c}{\sc Table 2: FOC Apparent Ionic Column Densities} 
\\[0.2cm]
\hline
\hline
%
\multicolumn{1}{c}{Ion}
&\multicolumn{1}{c}{Transition (\AA\/)}
&\multicolumn{1}{c}{$\log(\rm{N_{ion}})$}
\\
\multicolumn{1}{c}{(1)} & \multicolumn{1}{c}{(2)} & 
\multicolumn{1}{c}{(3)}  
\\[0.05cm]
\hline
\aliii & 1857 & 14.0$\pm0.15$   \\ 
\civ & 1549 &  15.7$\pm0.1$ \\ 
\siiv & 1397 & 15.4$\pm0.1$   \\
\nv & 1240 &   15.2$\pm0.1$   
\\[0.01cm]
\hline
\end{tabular}
\end{center}
\end{table}

\pagebreak

\section{IONIZATION EQUILIBRIUM AND ABUNDANCES}

In \S~2 we established that the \hei\ and \feii\ absorption featured
are not saturated, therefore their apparent column density
measurements are the actual column densities of the absorber and not
just lower limits.  The importance of this verification can hardly be
overstated.  Unless we have a direct evidence that a specific
absorption trough in a quasar's outflow is not saturated, we must
assume that it is, since in most cases where saturation diagnostics exist,
we find that the troughs are indeed saturated (Arav 1997; Telfer et
al. 1998; Arav et al. 1999a; Churchill et al. 1999; Arav et al. 1999b;
De Kool et al. 2000).  In these cases, using the apparent column
densities as real ones  undermine the ionization equilibrium
and abundances results inferred from ionization models, since the
output of these models are the real column densities.

\subsection{Optically Thin Models}

We begin by examining photoionization models that are optically thin
in the Lyman limit ($\vy{\tau}{LL}\ll 1$).  The output we are most
interested in are the ionic column densities ($\rm{N_{ion}}$), These
depend strongly on the input ionization parameter ($U$), total
hydrogen column density ($\rm{N_H}$) and metalicity.  To a lesser
extent $\rm{N_{ion}}$ also depend on the shape of the incident
continuum and are relatively insensitive to variation of $\vy{n}{H}$
in the range $ 10^5$ cm$^{-3}$ (our lower limit based on the \feii\
troughs) to $ 10^{10}$ cm$^{-3}$ (the estimated density in the broad
emission line region). With all other parameters fixed, $\rm{N_{ion}}$
depends linearly on $\rm{N_H}$, the atoms become more ionized with the
increase of $U$ and a higher metalicity correlates linearly with
higher $\rm{N_{ion}}$ for the metals.  In all our models we assume
solar metalicity. The elimination of this degree of freedom naturally
tightens the constraints we derive for the ionization equilibrium.  As
we show below, models based on solar metalicity can satisfactorily
reproduce the observed $\rm{N_{ion}}$.  For the incident spectrum we
used AGN continua given in the photoionization package CLOUDY (Ferland
et al. 1996): Table AGN, which is essentially the Mathews and Ferland
spectrum (Mathews and Ferland 1987) and two variants of a
superposition of a black body and various power laws.  We found that
the result are only moderately dependent on which of the three input
continua is used (see \S~4.3) and therefore we concentrate on results obtained using the
Table AGN continuum.

\begin{figure}
\centerline{\psfig{file=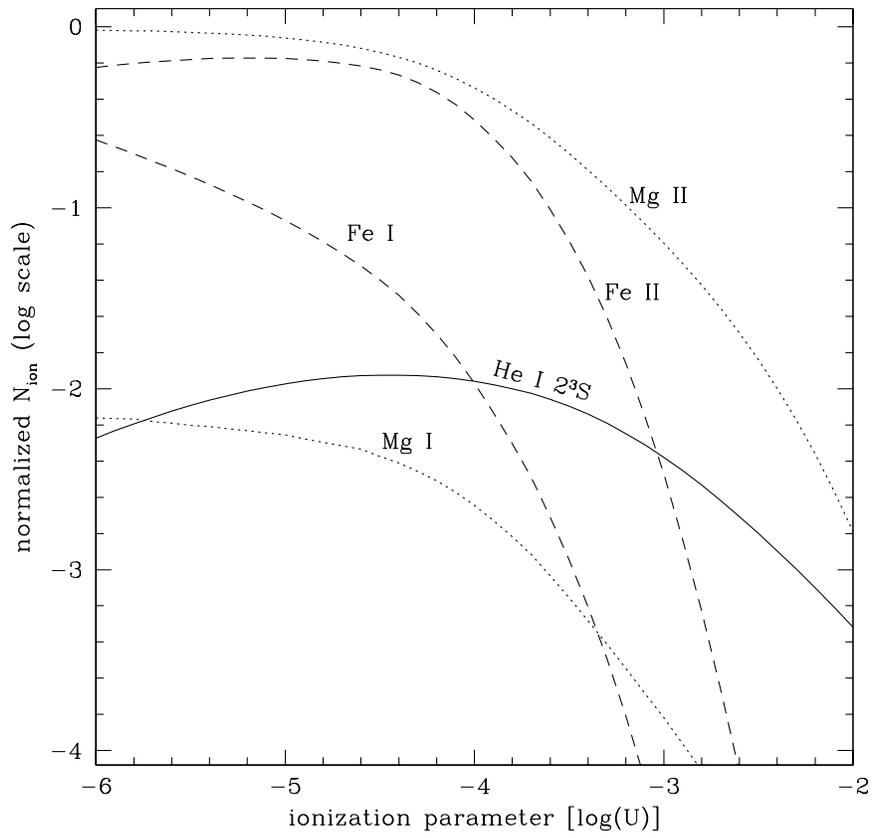,height=12.0cm,width=12.0cm}}
\caption{Optically thin ionization models ($\vy{\tau}{LL}\ll 1$), using the 
``table agn'' continuum with $\vy{n}{H}=10^8$ cm$^{-3}$. 
The ionic column densities are normalized to the total magnesium
column density.  
 }
\label{cloudy_u}
\end{figure}

 Figure \ref{cloudy_u} shows the relative $\rm{N_{ion}}$ as a function
of $U$ for most of the ions we detect in the HIRES spectrum.  The most
important constraint arise from comparing the prediction and
measurement for the ratio $\rm{N_{\feii}}/\rm{N_{\hei^*}}$ (where
\hei$^*$ designate \hei\ in the 2$^3$S metastable level).  When
\feii\ is the dominant iron ion, we expect
$\rm{N_{\feii}}/\rm{N_{\hei^*}}$ to be between 30--100 since the
solar abundance of iron is $3.2\times10^{-5}$ relative to hydrogen whereas
that of the \hei$^*$ is only $\sim6\times10^{-7}$ (based on
eq. \ref{eq:he_ratio}, assuming most helium is in \heii\ and using
helium abundance of 10\% relative to hydrogen; we note that the
temperature of the models ranges from 8000 K to 20,000 for $log(U)$
between --6 and --2, respectively). Since the observed
$\rm{N_{\hei^*}}$ is somewhat larger than $\rm{N_{\feii}}$ we conclude
that \feii\ cannot be the dominant iron ion.  From figure
\ref{cloudy_u} we deduce that the observed
$\rm{N_{\feii}}/\rm{N_{\hei^*}}$ exclude models with $log(U)<-3.5$.
Furthermore, the considerable difference in the slopes of
$\rm{N_{\feii}}$ and $\rm{N_{\hei^*}}$ allows for only a  narrow range
around $log(U)=-3$ for acceptable models.

What about the other observed ions?  The magnesium column densities fits
very well into the above picture.  Comparing the measurements in table
1 to the models in figure \ref{cloudy_u} we note that the observed
$\rm{N_{\mgi}}/\rm{N_{\hei^*}}$ can be reproduced by the models only in a
narrow range of $log(U)\gtorder-3$. The modeled $\rm{N_{\mgii}}$ are
consistent with this picture since the measurements are only lower
limits (see \S~2.3).  For \caii\ we also get consistent results 
and the upper limit for $\rm{N_{\fei}}/\rm{N_{\feii}}$ is readily 
satisfied as long as $log(U)>-5$, while the upper limit for 
$\rm{N_{\fei}}/\rm{N_{\hei^*}}$ necessitates $log(U)>-3.5$

\subsection{Models with a Hydrogen Ionization Front}
Although optically thin models can produce all the observed 
$\rm{N_{ion}}$ ratios, they have difficulties in explaining some of 
the  measured column densities themselves, especially $\rm{N_{\hei^*}}$. 
For $log(U)=-3$, figure  \ref{cloudy_u} shows that 
$\rm{N_{Mg}}/\rm{N_{\hei^*}}\simeq250$, or using solar abundances
$\rm{N_{H}}/\rm{N_{\hei^*}}\simeq7\times10^6$. Therefore, in order to produce
the observed $\rm{N_{\hei^*}}=6\times10^{13}$ cm$^{-2}$ we need a total 
$\rm{N_{H}}=4\times10^{20}$ cm$^{-2}$.  This amount of $\rm{N_{H}}$ produces a 
strong hydrogen ionization front (i.e., a region where the dominant
hydrogen ion shifts from \hii\ to \hi)
with $\vy{\tau}{LL}> 1000$ for $log(U)=-3$,
thus invalidating our optically thin assumption. Moreover, models with 
such a thick hydrogen ionization front cannot produce the observed 
$\rm{N_{ion}}$ either.  For the specific example above, the predicted 
$\rm{N_{\feii}}/\rm{N_{\hei^*}}$ is more than 100 times the observed one.

\begin{figure}
\centerline{\psfig{file=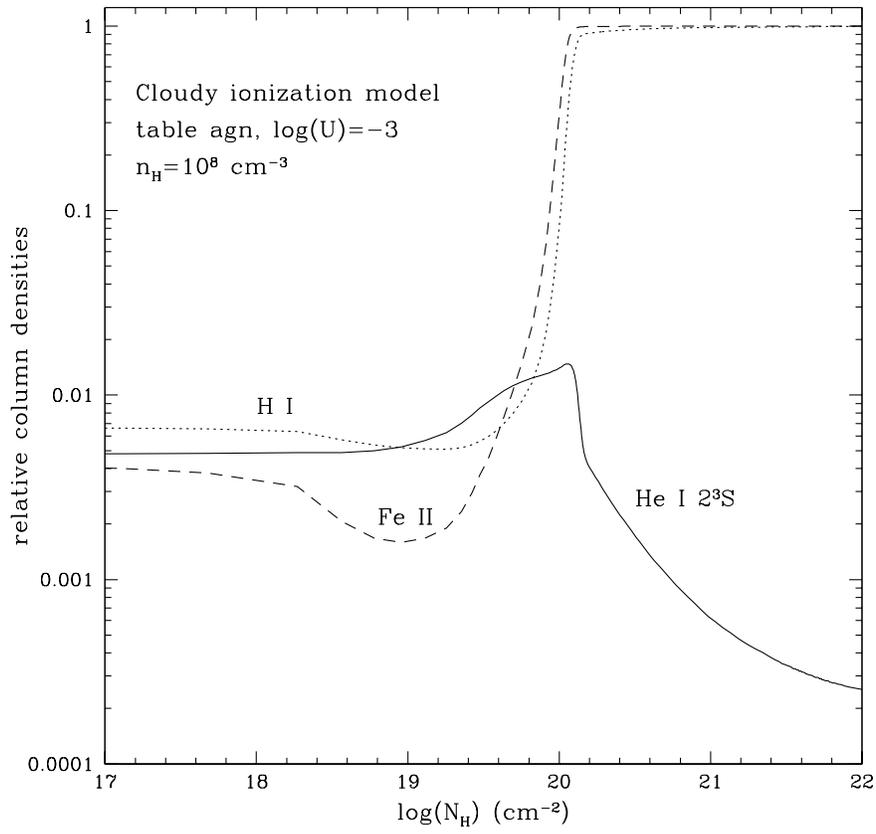,height=12.0cm,width=12.0cm}}
\caption{
Photoionization model with a strong hydrogen ionization front.
The \feii\ and \hei\ curves are normalized to the total iron column density
(assuming solar abundances).  
The \hi\ curve is normalized to the total $\rm{N_{H}}$.}
\label{cloudy_thick}
\end{figure}

Figure \ref{cloudy_thick} illustrates the situation for models with a
hydrogen ionization front.  Prior to the development of the hydrogen
ionization front, the relative column densities are similar to the
ones in the optically thin models.  We chose to present a model with
$log(U)=-3$ since the predicted ratio $\rm{N_{\feii}}/\rm{N_{\hei^*}}$
agreed with the observed one in the optically thin part.  Behind the
front the situation changes radically. The appearance of lines from
the \hei\ 2$^3$S metastable level requires a significant fraction of
\heii\ in the gas.  However, in close proximity to the \hi\ ionization
front we find a \hei\ ionization front making \hei\ the dominant
helium species.  Therefore, the fraction of \heii\ behind a hydrogen
ionization front decreases sharply and with it that of \hei$^*$ (see
eq.(\ref{eq:he_ratio})).  We note that the increase in the relative
fraction of \hei$^*$ in the vicinity of the front is mainly due to the
helium ionization transition.  For optically thin models with
$log(U)=-3$ the more abundant helium ion is \heiii\ (although only by
a factor 1.4 compared to \heii). As the equilibrium shifts towards
\hei, at some point \heii\ becomes most abundant, which causes the
abundance peak for the \hei$^*$.  As explained above, at a higher
column density \hei\ become the dominant species accompanied by a
sharp decline in the \heii\ fraction and hence in that of \hei$^*$.

For the \feii\ lines the situation is quite the opposite.  A hydrogen
ionization front protects the \feii\ ions from photoionization and
thus making it the dominant species.  As a result, almost all the iron
behind the front is in the form of \feii.  In the example given in
figure \ref{cloudy_thick} the relative fraction of \feii\ increases by
a factor of 300 across the front, whereas the fraction of \heii\ and
with it that of \hei$^*$ decreases sharply behind the front.
Regardless to the $U$ value of the incident spectrum, a strong
hydrogen ionization front always causes \feii\ to be the dominant iron
species behind the front.  Since even under the most favorable
conditions $\rm{N_{Fe}}/\rm{N_{\hei^*}}\gg 10$ (based on
eq. [\ref{eq:he_ratio}], assuming solar abundances and the allowed
temperature range discussed in \S~2.6), we conclude that models with a
 hydrogen ionization front  cannot
reproduce the observed \feii\ and \hei$^*$ column densities simultaneously.



\subsection{Realistic Model Fits}

From the analysis above we concluded that optically thin models have 
difficulties in reproducing the observed column
densities while models with a hydrogen ionization front fail completely.
  The obvious thing to try next  are models with an
intermediate \hi\ optical depth ($\vy{\tau}{LL}$ of a few).  We aim to
find models that can yield the observed column densities of component
$e$ (see Figs \ref{hires_he1} and \ref{hires_fe2}, and table 1) for
which we have the most accurate measurements.  In doing so we put most
weight on the \hei\ and \feii\ results since we have direct evidence
that the column densities we measure for these two ions are real ones
and not lower limits.  It is highly probable that the \mgi\
and \caii\ are also not saturated (due to similarities with the 
\feii\ and \hei\ features, as well as their moderate depth),
however we do not have direct diagnostics to confirm that.

\begin{figure}
\centerline{\psfig{file=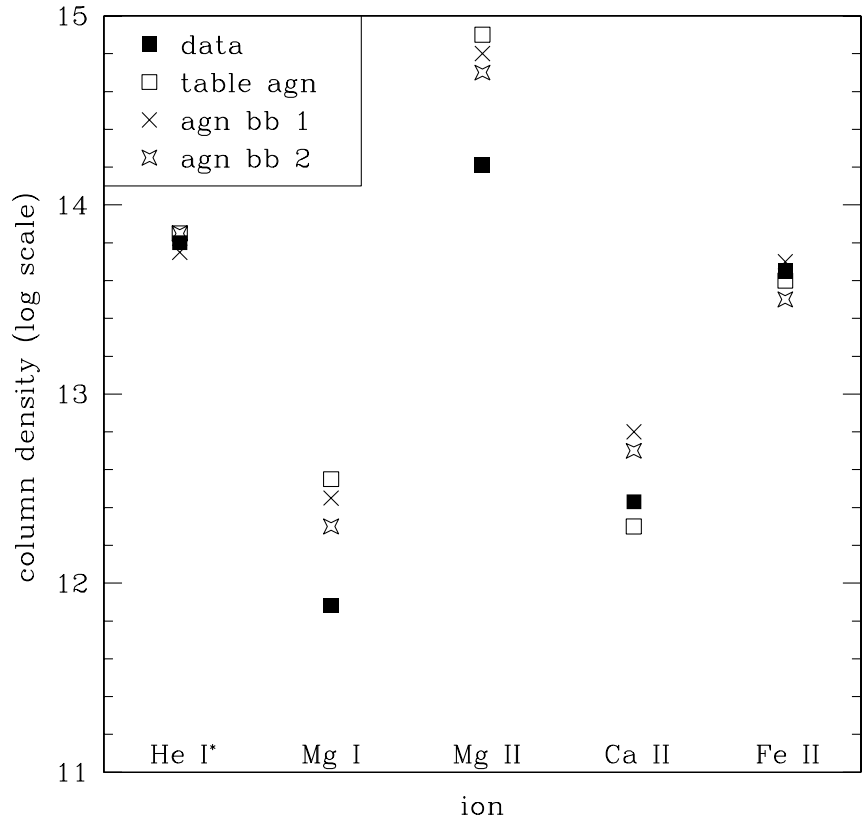,height=12.0cm,width=12.0cm}}
\caption{ Comparison between the observed column densities and simulated ones from three different 
models.  See Table 3 for models' details.}
\label{ionization_models}
\end{figure}

In figure \ref{ionization_models} we show the column density results
from three models as well as the observed ones.  In order to keep the
models simple  we used two AGN continua given in the
photoionization package CLOUDY (Ferland et al 1996; A full description
is given in the ``Hazy'' document which describes the code and is
available at: http://nimbus.pa.uky.edu/cloudy/cloudy\_94.htm).  The
``Table AGN'' model is fully determined by the package and is
described in detail in Mathews and Ferland (1987).  For the model
which is a superposition of a black body ``big bump'' and various
power laws, we used two settings: 1) the default parameter choices
given in the Hazy document , T=150,000 K, $\alpha_{ox}=-1.4$,
$\alpha_{uv}=-0.5$, $\alpha_{x}=-1$; 2) T=100,000 K, $\alpha_{ox}=-2$,
$\alpha_{uv}=-0.5$, $\alpha_{x}=-1$, which gives a somewhat lower
temperature across the cloud. The average temperature for each model
is given in Table 3, where the difference in temperature across each
model is less than 15\%.  As noted above, the column densities are
largely insensitive to variation of $\vy{n}{H}$ in the range $10^5$
cm$^{-3}$ to $10^{10}$ cm$^{-3}$. For the presented models we chose
$\vy{n}{H}$ value close to our \feii-inferred lower limit, since they
gave somewhat lower temperatures, in better agreement with the \feii\
temperature constraints.  The Table AGN and big bump models give
excellent fits for the \feii\ and \hei\ column densities.  The models
over-predict $\rm{N_{\mgi}}$ by about a factor of three and give a
reasonable fit for $\rm{N_{\caii}}$.  Although we did not include
aluminum in figure \ref{ionization_models} (since we do not have
high-resolution data for it), we note that these models also reproduce
the apparent $\rm{N_{\aliii}}$ (see table 2) to within a factor of
two. The low abundance of aluminum ($[Al/H]_{\odot}=3\times10^{-6})$
coupled with the inferred $\rm{N_{H}}$ in the object, suggests a low
level of saturation (if any) in the \aliii\ absorption, thus implying
 that the apparent $\rm{N_{\aliii}}$ is a reasonable approximation
for the actual one.

 How does the presence of other flow components affect our ionization
analysis?  Although the measurements of components $a-d$ are not as
accurate, it seems that their column density ratios, and therefore
their ionization equilibrium, are similar to that of component $e$.
This situation argues in favor of models with with lower
$\vy{\tau}{LL}$.  A moderately strong $\vy{\tau}{LL}$ ($\gtorder4$) in
a given component will strongly attenuate the incident ionizing
spectrum seen by  components further from the central source. 
As a result, their
ionization equilibrium will be different.  If component $e$ is the
furthest away from the source and is the only optically thick one, we
still need fine tuning to have similar column density ratios in
optically thin and optically thick slabs.  The higher 
$\vy{\tau}{LL}(e)$ is, the more difficult it is to explain the similar
ionization equilibrium in the other components.  Based on these
arguments we prefer models that fits the data with the smallest
$\vy{\tau}{LL}(e)$.  This is best achieved by choosing incident
continuum that yields low temperature.  As evident from equation
(\ref{eq:he_ratio}), with all else equal, we get higher concentration of
\hei\ in the 2$^3$S level at lower temperatures.  A smaller
$\rm{N_{H}}$ is then needed for a given $\rm{N_{\hei^*}}$, which leads
to smaller $\vy{\tau}{LL}$.  Additional support for lower temperature
models come from the \feii\ inferred T$<10,000$~K (see \S~2.6).



\pagebreak

\begin{table}[htb]
\begin{center}
\begin{tabular}{lccccc}
\multicolumn{3}{c}{\sc Table 3: Photoionization Models } 
\\[0.2cm]
\hline
\hline
%
\multicolumn{1}{c}{model}
&\multicolumn{1}{c}{$\log(\rm{N_{ion}})$}
&\multicolumn{1}{c}{$\log(U)$}
&\multicolumn{1}{c}{$\vy{n}{H}$}
&\multicolumn{1}{c}{$\vy{\tau}{LL}$}
&\multicolumn{1}{c}{$\overline{\rm{T}}$}
\\
\multicolumn{1}{c}{} 
& \multicolumn{1}{c}{(cm$^{-2}$)} 
& \multicolumn{1}{c}{}  
& \multicolumn{1}{c}{(cm$^{-3}$)}  
& \multicolumn{1}{c}{}  
& \multicolumn{1}{c}{($10^3$ K)}  
\\[0.05cm]
\hline
   table agn    & 20.2 & -2.7 & $10^6$ & 6.5 & 12 \\ 
   agn bb 1	& 20.0 & -2.8 & $10^5$ & 3.6 & 10 \\
   agn bb 2 	& 19.9 & -2.7 & $10^5$ & 1.5 & 8 
\\[0.01cm]
\hline
\end{tabular}
\end{center}
\end{table}

\subsection{Conditions Needed for Detecting \mgi\ Absorption}  

  In spite of their high S/N, our low-resolution observations do not
reveal absorption from \mgi\ $\lambda$2853.  The reason for this is
the small equivalent width of the \mgi\ features, which in our case
can only be detected with a combination of high S/N - high spectral
resolution data.  As we discuss in \S~2 the detection of \mgi\ puts
important constraints on photoionization models.  It is therefore
important to observe \mgii\ BALQSOs at high spectral resolution in
order to determine the existance of \mgi\ in the intrinsic absorber.

The ionization models account for the weakness of the \mgi\ features 
compared to
those of \mgii.  For material optically thin at the Lyman edge, figure
\ref{cloudy_u} shows that for a typical AGN spectrum
$\rm{N_{\mgii}}/\rm{N_{\mgi}}\gtorder100$ for $-6<\log(U)<-3$; this
result also holds up to  $\log(U)=-8$.
Thus, for $-8<\log(U)<-3$ the dominant magnesium ion in absorption is
\mgii\ with \mgi\ showing roughly 1\% of it's column density.  That
translates to a factor of 30 difference between the optical depths of
\mgii\ $\lambda$2796 and \mgi\ $\lambda$2853.  If the \mgii\ absorber
is highly saturated we may detect \mgi\ absorption even in
low-resolution data, as is the case in QSO~1044+3656 (de Kool et
al. 2000). Otherwise, only high resolution data can reveal its
existance in the deepest parts of the \mgii\ troughs, which is the
case in QSO~2359--1241.  In order to detect \mgii\ at
$\log(U)\gtorder-2$ a hydrogen ionization front is needed (Voit,
Weymann \& Korista 1993) to keep the relative fraction of \mgii\ ions
high enough.  But in that case, we should not expect to see \mgi\
absorption features. For example, a model using table AGN continuum,
$\log(U)=-2, \vy{N}{H}=10^{22}$ cm$^{-2}$ (a combination which
produces a thick hydrogen ionization front) and $\vy{n}{H}=10^{8}$
cm$^{-3}$, predicts that behind the front 99\% of the magnesium is in
the form of \mgii\ while only $2\times10^{-4}$ is in \mgi.  A good
observational example is QSO~0059--2735, where \mgii\ is detected,
\mgi\ is not (Wampler, Chugai \& Petitjean 1995) and a black Lyman
limit is observed (Turnshek et al. 1996).

\subsection{Relationship Between the High and 
Low Ionization Absorbers}

What is the relationship between the low-ionization absorption (\mgii,
\feii, \mgi...) and the high-ionization absorption (\civ, \siiv,
\nv...) that are seen in the spectrum of low-ionization BALQSOs?
Voit, Weymann \& Korista (1993) noted that the low-ionization
absorption tends to concentrate at low ejection velocity compared to
the high ionization absorption, where the latter normally encompass
the velocity range of the former and extends to a much higher
velocity.  The best examples for that behavior are QSO~0059--2735 and
QSO~1232+1325.  However, this is not always the case.  QSO~0932+5010
shows strong \mgii\ absorption in its high velocity trough (data
courtesy of Ray Weymann and Kirk Korista) and in QSO~2359--1241 we
observe a high velocity \mgii\ trough at --5000 \kms.  As discussed in
\S~3, our HST UV prism-spectroscopy data show a wide BAL trough in
both \siiv\ and \civ\ (FWHM $\sim8000$ \kms), which encompasses both
the high and low velocity troughs seen in \mgii. This relationship
strongly suggests that the low-ionization absorption seen in
QSO~2359--1241 is indeed physically connected to the BAL flow seen in
the UV lines.  It appears that a simple picture where the low
ionization outflow is confined to low velocity does not hold.  A more
elaborate model, perhaps including ionization stratification (Arav et
al. 1999a), is called for.  This picture is strengthen by the
observation that in QSO~0059--2735 there is a clear detection of \ovi\
$\lambda1034$ BAL (Turnshek et al. 1996).  It is very difficult to
construct a single-zone ionization model where significant optical
depth arises from both \mgii\ and \ovi.  We point out that ionization
models with large local density gradient give natural explanation for
such occurrence (Arav 1996; Arav et al. 1999a) since they allow for
material with large variation in ionization parameter to exist at
close proximity.

\section{SUMMARY}

The spectrum of QSO~2359--1241 contains powerful diagnostics for the
state of it's outflow.  In particular we emphasize the importance of
the \hei\ lines from the 2$^3$S meta-stable level.  Under optimal
conditions, the abundance of this level is roughly 30 times lower than
the abundances of iron and magnesium.  Detecting a somewhat higher
column density of the \hei\ 2$^3$S meta-stable level than that of
\feii\ indicates that 95--99\% of the the iron is in higher ionization
stages and allows for tight constraints on the ionization parameter.
The three well separated \hei\ lines also allow for an excellent
saturation/covering-factor analysis, since their oscillator strength
differ by a factor of six.  With the combined constraints available
from the \feii\ and \hei\ lines we are able to tightly constrain the
ionization equilibrium in the flow; reproduce the observed column
densities without invoking departure from solar abundances; and
exclude a hydrogen ionization front in this outflow.

A strong connection between the low ionization absorber and the BAL
phenomenon is evident in the HST FOC data.  High ionization BALs (from
\civ, \siv\ and \nv) are detected in the HST data and these encompass
and expand the velocity range seen in the low ionization species.  It
will be very valuable to obtain better HST spectroscopy of the high
ionization lines, as well as ground observations of the
\aliii~$\lambda$1857 absorption.  These will allow to study the
connection between the high and low ionization absorbers in greater
detail, and will yield additional constraints on the low-ionization
absorber through the \aliii~$\lambda$1857 and \alii~$\lambda$1670
lines.

\section*{ACKNOWLEDGMENTS}
We thank Kirk Korista, Martijn de Kool and the referee for numerous
valuable suggestions.  We acknowledge support from NASA HST grant
GO-06350, NSF grant AST-9802791 and STScI.  Part of
this work was performed under the auspices of the US Department of
Energy by Lawrence Livermore National Laboratory under Contract
W-7405-Eng-48.

\end{document}